\documentclass[]{spie}  

\usepackage{amsmath,amsfonts,amssymb}
\usepackage{graphicx}
\usepackage[colorlinks=true, allcolors=blue]{hyperref}
\usepackage{listings}
\usepackage{color}

\usepackage{wrapfig}
\usepackage{lscape}
\usepackage{rotating}
\usepackage{epstopdf}
\usepackage{astro_bib_macro} 

\definecolor{dkgreen}{rgb}{0,0.6,0}
\definecolor{gray}{rgb}{0.5,0.5,0.5}
\definecolor{mauve}{rgb}{0.58,0,0.82}

\title{High-contrast Imager for Complex Aperture Telescopes (HiCAT): 6. Software Control Infrastructure and Calibration}

\author{Christopher Moriarty, Keira Brooks, R\'{e}mi Soummer, Marshall Perrin, Thomas Comeau, Gregory Brady, Rob Gontrum, Peter Petrone
\skiplinehalf
Space Telescope Science Institute, 3700 San Martin Drive, Baltimore, MD 21218, USA
}

\authorinfo{Further author information, send correspondence to Christopher Moriarty: E-mail: moriarty@stsci.edu}

\begin{document} 
\maketitle

\begin{abstract}
High contrast imaging using coronagraphy is one of the main avenues to enable the search for life on extrasolar Earth analogs. The HiCAT testbed aims to demonstrate coronagraphy and wavefront control for segmented on-axis space telescopes as envisioned for a future large UV optical IR mission (LUVOIR). Our software infrastructure enables 24/7 automated operation of high-contrast imaging experiments while monitoring for safe operating parameters, along with graceful shutdown processes for unsafe conditions or unexpected errors. The infrastructure also includes a calibration suite that can run nightly to catch regressions and track optical performance changes over time, and a testbed simulator to support software development and testing, as well as optical modeling necessary for high-contrast algorithms. This paper presents a design and implementation of testbed control software to leverage continuous integration whether the testbed is available or not.

\end{abstract}


\section{INTRODUCTION}
\label{sec:INTRODUCTION}

High contrast imaging on telescopes with a segmented primary mirror, secondary mirror, and secondary mirror supports is a challenge that must be solved to enable the search for life on terrestial analogs around nearby stars. The High contrast imaging for Complex Aperture Telescopes (HiCAT) aims to reach several technological advances in the area of coronagraphic design, wavefront control and the overall system integration for a simulated on-axis segmented space telescope \cite{2016SPIE.9904E..3CL, 2017SPIE10562E..2ZL, ndiaye2015AAS,ndiaye2014SPIE_hicat2, ndiaye2013SPIE}.

One of the primary goals of HiCAT is to test an Apodized Pupil Lyot Coronagraph (APLC) solution for on-axis segmented telescope architectures \cite{ndiaye2014SPIE_aplc}. This partially obscured aperture scatters light in a complex, though predictable way, which is mitigated by the APLC optimization. In addition, wavefront control using deformable mirror is able to create a ``dark zone'' by nulling speckles created by the scattered light in the expected location of possible companions to the star being observed. Because of the multiple components and masks (pupil mask, segmented aperture deformable mirror, apodizer, lyot stop, etc.) the testbed requires flexibility in hardware to swap between multiple configurations (e.g. full circular aperture vs. segmented aperture, apodized vs. non-apodized pupil, etc.). These sensitive and critical optics are often in hard-to-reach places on the testbed. Performing these swaps involves many steps that can affect the alignment and other qualities in the setup (more information given by Soummer in these proceedings\cite{soummer2018}). 
While this presents a set of challenges with regard to alignment and testbed architecture this also creates challenges in the implementation and running of this testbed from a software standpoint (including re-calibration after hardware configuration swaps). We designed the software infrastructure with these challenges in mind. Despite having to interface to hardware components from a multitude of different vendors, programming languages and environments, the complexities were abstracted with an object-oriented architecture so development of experiments and algorithms can be done without the syntactical clutter of complex hardware communication.

The main HiCAT testbed construction was completed several years ago and since becoming operational, has had a multi-language LabVIEW-based control system. About a year and a half ago, the HiCAT team decided to move away from such a system and toward a unified control infrastructure, written in one language. This transition included using version control, bug tracking, and a feature branching workflow in order to incorporate multiple developers and a flexible process for managing the new team and evolving goals.  The new software infrastructure would be object-oriented and therefore more easily extensible when hardware changes occur, such as swapping cameras or apodizers. Using version control and feature branching provides reliable code history and a structured release process. The transition to software engineering best practices has enabled the team to adapt to some new challenges when working with an ever-changing and frequently unavailable testbed.

Continuous Integration (CI) is a software engineering practice that strives to reduce development and integration time by regularly running builds and tests from a shared code repository. Applying CI to a software project has become standard practice, and tools such as Jenkins\cite{jenkins} are freely available and widely adopted.  This paper lays out an implementation of a software control system for a high contrast imaging testbed, HiCAT, which was designed from the ground up to make use of CI for both software and hardware. Not only will applying CI bring benefits of better integrated software, but having nightly testing running on the HiCAT testbed will allow for trending and analysis of how the testbed changes over time. Currently, we can run the equivalent of a nightly build of the testbed in the form of focus calibration, subarray correction, plate scale calculation, and specifically in the case of the main science goal: contrast achievements. 

Laying the foundation to apply and maintain a robust CI infrastructure was done while having to meet critical deadlines outside of the scope of software tasks. The updated software design first consolidated functionality under one programming language to reduce the complexities between interfaces and ramp-up time for new developers.  Another important factor for a CI system with a hardware component is the ability to run builds and tests even if the hardware is unavailable.  To account for this, the software interfaces developed for hardware control make use of an abstract class that standardizes the syntax and instills proper resource management.  With an object-oriented architecture, the team can create simulated versions of each of the hardware components and in effect run experiments offline using the same operational code.

Above all else, safe hardware control is paramount to the success of the project.  Several components on the testbed are sensitive and critical to the design, and if they were to fail it could have significant impact. The hardware control architecture takes into account that some bugs and errors are out of the control of the developers. The hardware interface design contains a smart, fool-proof resource control mechanism that properly closes connections to hardware, always leaving them in a safe state.  Along with resource management, an automated safety monitoring was introduced into the architecture.  Every experiment written for HiCAT inherits these features through an elegant object-oriented design.

\section{Unifying and Versioning the Software}
\label{sec:Transistion to Python}

The initial control system for HiCAT consisted of several different scripts, notebooks, and executables written in multiple languages, all strung together with LabVIEW. LabVIEW offers a graphical user interface, and allows non-developers to design and implement control systems.  This is useful for many scenarios, but struggles to scale for a large complex instrument like HiCAT. We decided to consolidate most, if not all, of the code under one language or technology, and move away from LabVIEW. This would foster a more maintainable software system, and result in the ability to design a more intuitive and user-friendly system. 

\subsection{Migration to Python}
The astronomy community has benefited from huge strides in the efficiency and science capabilities of Python and its popular libraries.  In addition to being open source and a widely used programming language, Python promotes a low barrier to entry for new developers. With this in mind, all existing HiCAT code was migrated to Python, making use of well-supported libraries such as NumPy\cite{numpy} and SciPy\cite{scipy} to replace components written in MATLAB\cite{matlab_to_python} and Jupyter Notebook to replace the Wolfram Mathematica notebooks. 

The first hurdle of the migration to Python was that the majority of software development kits (SDK) and application programming interfaces (API) for the hardware are best supported by a Microsoft-based language (C++/C\#). A few hardware vendors do provide a Python SDK, although in at least one case, it was only for Python 2.7.  According to PEP 373 (the Python 2.7 Release Schedule), the end-of-life for Python 2.7 is set for 2020.  To prepare for this, the advice provided by \href{python-future.org}{python-future.org} was followed, which demonstrates how to use a few import statements to facilitate the ability write Python 3.6 code that also runs as expected in Python 2.7. Once vendors release a Python 3.6 compatible SDK, it is possible to smoothly and completely transition the environment towards Python 3.6.  

For those hardware vendors that do not provide a Python SDK, libraries such as \verb|cdll| and other low-level communication packages allowed successful development of interfaces to each hardware component. This has given rise to another goal: to create a library of robust hardware control interfaces in Python to share with other projects and testbeds.  To enable this, we made the HiCAT code into a proper Python package which can be installed using pip, a common package manager included in most Python environments. Our collaborators can now easily access the code, and when installing with pip, automatically gather and install the necessary dependencies. 

\subsection{Version Control}
Tracking code changes to the software is absolutely critical for troubleshooting problems with the testbed. A testbed left alone can morph, be it from temperature changes or warping of the table, therefore it is essential to be able to easily roll back software to a known-to-work state. 

Along with choosing Python, the lab also adopted Git for version control and Jira for issue tracking. Git is a distributed version control system (DVCS) where every developer's working copy of the code is a repository that can contain the full history of changes. Prior to using version control, the lab's code was simply stored on a network file share, which made it risky to make changes, and onerous to recover previous working copies. Having the code on a shared network drive inside of the STScI firewall also made it difficult to allow external collaborators to contribute. The hicat-package repository is now hosted on GitHub, under the STScI account, and external collaborators can easily be invited and control permissions given as needed. 

In order to provide a scalable development workflow as the team expands, we adopted a feature branching workflow. This workflow, GitFlow, does not add any new concepts or commands, instead it streamlines feature branching  by assigning roles to branches and defining how they should interact. Under this paradigm, all tickets correspond to a branch, and code is tested and reviewed before being merged into the "develop" branch, which is the main development branch for the current release.  GitFlow also provides a clean history of released versions, called the "master" branch, logically separating releases from development.  To track branches properly between Jira and GitFlow, branches always have a matching Jira issue. Branches only get merged once they have a chance to move through each state of a workflow created within Jira, which includes testing on the hardware with the latest version of the development branch.

\section{Hardware Interface Design}
\label{sec:Hardware Interface Design}

As previously mentioned, the HiCAT testbed has hardware components from many different vendors, and as such, it can be easy to end up with a litany of different software interfaces. It can also be tempting to use the example code bundled with the device, which often is in Matlab, C++, or LabVIEW.  While there is nothing wrong with those technologies independently, it is easy for a codebase to become spaghetti code if every device has its own unique syntax to communicate. Therefore, we decided to create a standardized layer for hardware control, also making scripting experiments much more intuitive; one need not understand the low-level communication layer in order to safely control the testbed.

Another motivation for the chosen hardware interface infrastructure is operating the hardware safely, even when unexpected errors or exceptions occur. This is done by abstracting the different types of hardware (camera, motor controller, etc), so they can be implemented as mock or simulated classes to support offline testing and ultimately enable CI.  The interfaces to hardware should be free of any HiCAT specific code.  Keeping the code generic enables sharing with other testbeds that have similar hardware components. This has recently been demonstrated with the hicat-package installation for the JWST Optical Simulation Testbed (JOST)\cite{laginja2018} to control several hardware components. We aim for our repository to be made public, and eventually open source to foster a community that extends the number of devices supported.    

An additional challenge is having to take into account not just the number of components on the testbed, but also the complexity of the components. One of the most difficult hardware components to accommodate on HiCAT are the deformable mirrors (DM). HiCAT has two Boston Micromachines kilo DMs (952 actuators - 34 across - in a circular aperture) which only have a one-way communication interface.  This means that the software can send commands to the DMs, but is unable to query for status. The DMs are also required to be under humidity control, and properly closed to ensure that a voltage is not applied when not in use. To mitigate the risk involved with resource management, Python has a construct called a ``context manager", which became an integral part of our design.

\subsection{Context Managers}

Similar to the concept of a try-catch, a context manager will run defined code even when errors or exceptions are thrown. When managing connections to hardware, context managers provide reliable and elegant mechanisms to ensure connections are always initialized, operated, and closed properly.

A class can be turned into a context manager by declaring and implementing two functions: \verb|enter| and \verb|exit|.  This allows the use of the Python \verb|with| keyword, which specifically makes calls to the aforementioned functions.  Just as the name suggests, the \verb|exit| function is called when the code falls out of context. This happens in the normal flow of the code, or it could happen because an error gets thrown; a model that is much preferred over a sequential operating mode, where you explicitly open and close the connection. In a sequential operating mode, if the user opens the connection and then experiences an unexpected error, the code may not properly close the connection. In the case of the one-way interface to the DM, this implies that there would still a be voltage applied to the actuators, potentially causing significant damage to a vital component in the system.

In the spirit of inheritance and object-oriented design, an \verb|Instrument| parent abstract class was created as a context manager and to provide two abstract functions (\verb|initialize|, and \verb|close|). These two functions will not need to be explicitly called, instead Python detects that a context manager is being used and handles the object in a special manner.  Each hardware type then also has an abstract class which defines the function stubs for any children of the same type, then the children of the hardware-specific types implement the actual hardware control.  This object-oriented approach allows developers to script experiments in such a way that never ties their code to any specific piece of hardware. Overall, the design instills proper resource management into every hardware interface that is developed. 

The code for the \verb|Instrument| abstract base class has three sections (see full code in Listing \ref{lis:instrument_class}):
\begin{enumerate}
\item Constructor
\item Context Manager implementation (\verb|__enter__()| and \verb|__exit__()|)
\item Abstract \verb|initialize()| and \verb|close()| methods
\end{enumerate}
The constructor accepts a parameter named \verb|config_id|, which is simply a string that acts as a key into the configuration file (config.ini). More importantly, it calls the \verb|initialize| abstract function, which would perform any required initialization to the hardware, and return a connection object. The object will get stored as the class attribute: \verb|instrument|. The \verb|enter| function returns a reference to the initialized object, mostly as a way to enable the \verb|with| keyword syntax.  The \verb|exit| function calls the abstract \verb|close| function, and once completed it sets the instrument attribute to \verb|None|.  Python does not create a new scope within the \verb|with| statement, so even though the object falls out of scope and calls its \verb|exit| function, it will still remain accessible, and therefore prone to accidental reuse. Now, instead of an unknown error getting thrown from a hardware-dependant API, the reference will not exist, immediately communicating to the developer that something has gone wrong.  
\pagebreak

\begin{lstlisting}[caption={The actual code for the \texttt{Instrument} class, which acts as the standardized way to implement an interface to a hardware device.  By inherited instrument, the child class becomes a context manager, and by implementing "initialize" and "close", the child class can becomes a reliable resource manager.},captionpos=b,label={lis:instrument_class}]
class Instrument(object):
    __metaclass__ = ABCMeta

    def __init__(self, config_id, *args, **kwargs):
        self.config_id = config_id
        self.instrument = self.initialize(self, *args, **kwargs)

    # Implementing context manager enter and exit functions.
    def __enter__(self):
        return self

    def __exit__(self, exception_type, exception_value, exception_traceback):
        self.close()

        # Set the instrument to None to ensure it doesn't get reused.
        self.instrument = None

    # Abstract Methods.
    @abstractmethod
    def initialize(self, *args, **kwargs):
        """Implementation should return the resource object provided by sdk/api."""

    @abstractmethod
    def close(self):
        """Implement this function to close and cleanup instrument connection."""
\end{lstlisting}

\begin{figure}
	\centering
	\includegraphics[width=\linewidth]{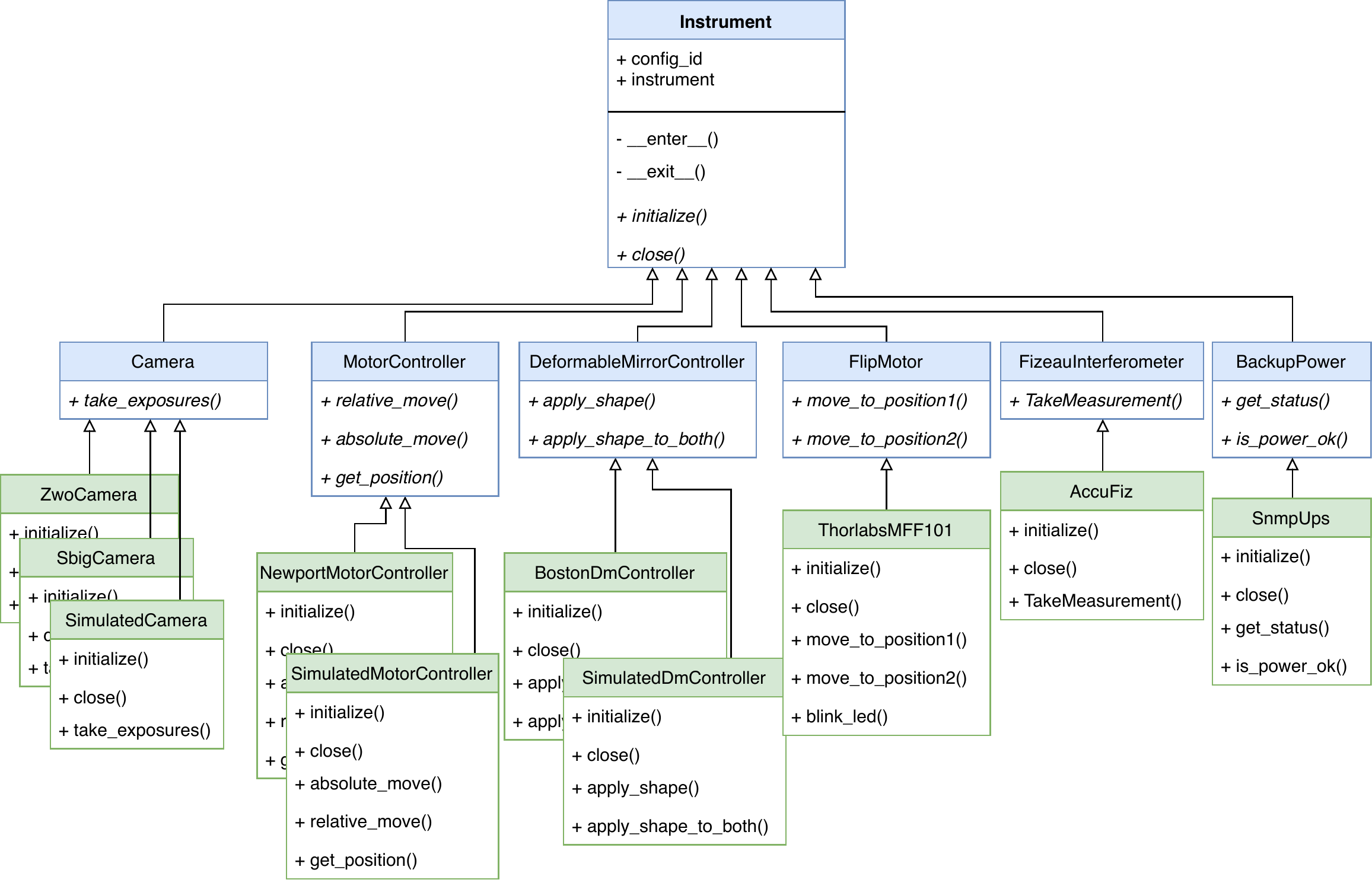} 
    \caption{Class diagram of the hardware control interface structure of the hicat-package. Shows how the the Instrument parent class is inherited and instills the context manager functionality in all children.}
    \label{fig:class_diagram}
\end{figure}

\section{Calibration Suite}
\label{sec:Calibration Suite}

A testbed such as HiCAT, that, by the nature of the project, will have major optical components exchanged, must also have an easy way to calibrate the testbed between all hardware changes. This includes mapping how the testbed changes over time, and checking to make sure the testbed is in a known state after a hardware-swap. 

In addition to being a way to return the testbed back to a predictable state, other high contrast imaging testbeds have found that components can degrade over time, and cause the PSF to change significantly. By mapping the state of the testbed over time, any changes that occur in the system are monitored and the source pinpointed. This trending will be available once the testbed is in a stable state and regular tests are put in place for any desired timescale. 

Calibration activities with HiCAT have been an integral part of the workflow process since the switch to Python and a re-mapping of the interface between software and hardware. Because of this, the calibration activities act as end-to-end regression tests for these interfaces. It is known what results to expect, and if a result differs wildly, we can catch the regression early and even revert back need be. If it is determined that the new results are correct, we have documented and mostly automated processes for re-calibrating the system. The calibration activity algorithms evolved to be more robust to changes in the hardware configuration by using the calibration activities for different testbed states (i.e. when an apodizer is added to the system and changes the expected speckle pattern, therefore changing how centering is performed). This can easily be used for regular health testing, even if no hardware changes have taken place. Running them regular creates logs of the results and enables trending of testbed stability.
\pagebreak

\begin{figure}
	\centering
	\includegraphics[width=\linewidth]{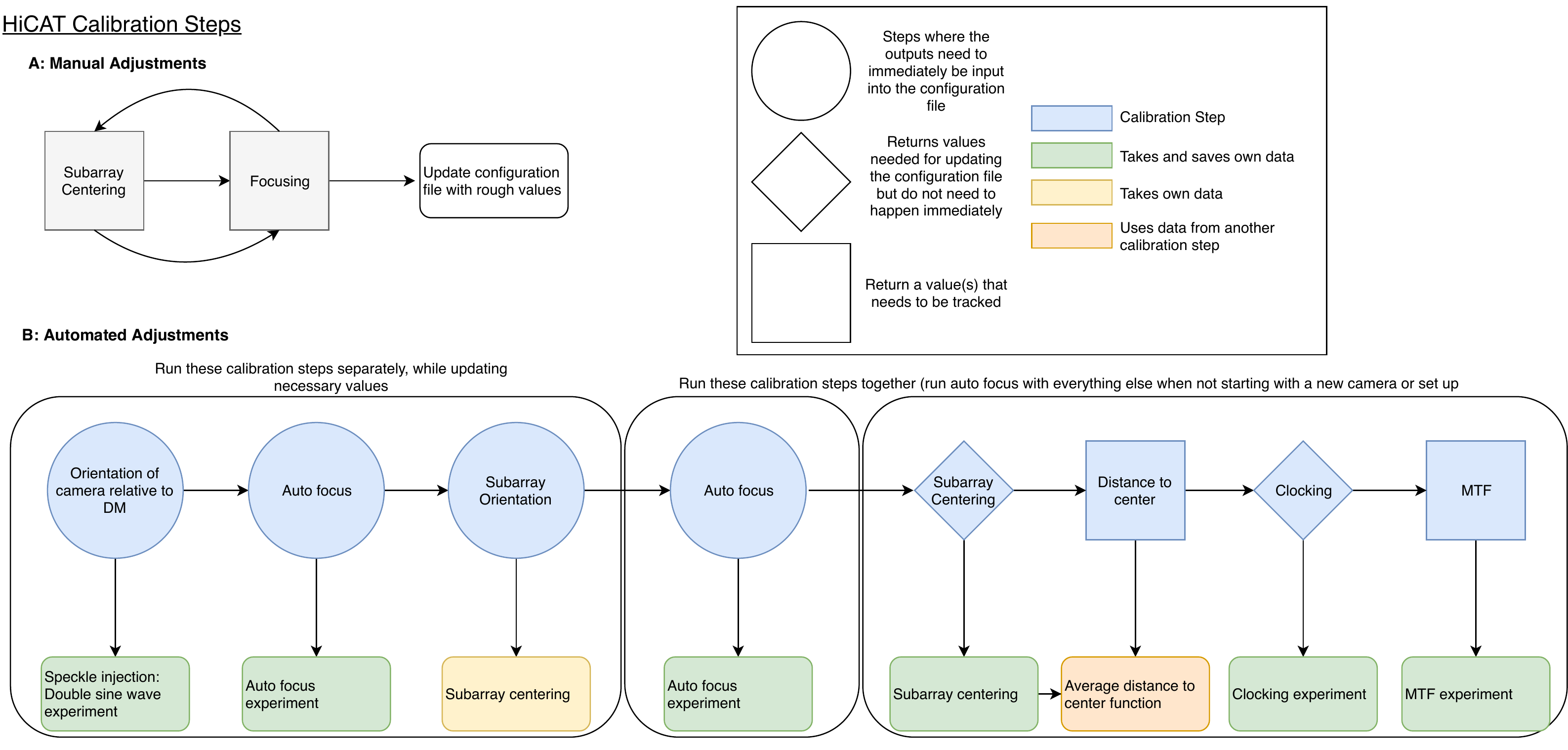} 
    \caption{Diagram shows the different calibration activities, and the flow and dependencies of how they work together.}
    \label{fig:calibration_steps}
\end{figure}

Currently, a small suite of activities that are used regularly (see also Figure \ref{fig:calibration_steps}):
\begin{itemize}
    \item Orientation of the camera 
    \item Clocking of the camera
    \item Location of image and changes in centering on the detector
    \item Camera position for best focus
    \item Best location of focal plane mask (FPM)
    \item Sampling/plate scale
\end{itemize}

While these calibration activities have been in place for approximately a year, due to regular changes to the testbed hardware, trending has not yet been possible, though results from the calibration activities have been saved. The number of activities will evolve as the testbed state become more stable and the preliminary list for the next wave of calibration activities to be created is as follows:
\begin{itemize}
    \item Laser linearity and flux stability
    \item Wavefront error correction with phase retrieval (see Brady\cite{brady2018} in these proceedings).
    \item Contrast calibration
\end{itemize}

With the addition of a high-fidelity simulator (see Section \ref{sec:Simulator}) expected errors can be added in the trending analysis to know when changes in the testbed are cause for concern.

\section{Real-time Safety Monitoring}
\label{sec:Safety Monitoring}

Python context managers and the \verb|Instrument| class provide standardized resource management for the hardware devices, although without automated safety monitoring it is impossible to run the testbed without human interaction. Specifically for HiCAT, the most crucial safety concerns are humidity and stability of power. Instead of only implementing these two, an architecture was created to incorporate any number of safety monitoring mechanisms. This real-time safety monitoring approach enables any code controlling HiCAT to automatically spin off a parallel process that checks for safety concerns, with the capability to safely shutdown the hardware if a safety concern is discovered. CI becomes much more practical if the software and hardware can be controlled autonomously.

\subsection{Parallel Processing in Python}
Conceptually, spinning off a separate thread to do safety monitoring is a simple idea. Hardware control should not be interrupted or paused by reading sensors or other safety checks, so parallel processing solves that.  Python, however, cannot guarantee that a thread will be run in parallel because of the Global Interpreter Lock\cite{gil} (GIL).  Therefore, it is necessary to make use of the multiprocessing library to ensure safety monitoring will not delay the experiment or vice versa.  With multiprocessing, a completely separate Python process is created and avoids the inherent drawbacks of multi-threading with the GIL. 

For Linux and Mac OSX operating systems (OS), Python processes can use the \verb|signal| library to send commands between processes, such as a signal interrupt, which is commonly used as a soft kill. Context managers are triggered by certain signals, although it's also possible to kill a process abruptly with the \verb|signal| library. In contrast, the Windows OS only supports a subset of available signals, none of which are ones that a context manager can catch.  Even worse, Python will allow a developer to write code that seems to be sending a signal interrupt, but the Windows OS interprets it as a hard kill and does not allow context managers to gracefully exit.  However, there are Windows OS-specific Python libraries called \verb|win32api| and \verb|win32con|, that can generate Windows events such as a CTRL+C event.  Context managers are able to handle such events gracefully, because they are interpreted as a \verb|KeyboardInterrupt|. HiCAT is controlled by a Windows PC, so these libraries were used and provided similar functionality, albeit a bit less elegant. Instead of sending a signal only to an individual process, a CTRL+C event is raised and any active context managers close properly before exiting completely. It should be noted that all processes running within the console created when Python was started would also receive this command. 

\section{Autonomous Operations}
\label{sec:Autonomous Operations}

Leveraging the parallel real-time safety monitoring design, a class called \verb|Experiment| was created to allows any subclass of it to inherit safety monitoring automatically.  The class is structured such that the developers writing new experiments need only implement a single abstract function called \verb|experiment()|, and the rest of the safety monitoring functionality gets instilled through inheritance. This allows any number of experiments to be created as objects and queued up. Queuing experiments allows the user to plan a list of tasks to accomplish overnight or on weekends, while ensuring safety. The queue shuts down if safety concerns are found, although continues if there is an unexpected error along the way. The logic for this is built into what is called a ``launcher", a concept in the infrastructure that enables the implementation of CI. Launchers are simple python scripts that allows the creation of a list of experiments to run be in an event-based operational paradigm. 

\subsection{Experiment Class}
The \verb|Experiment| class aligns well with the object-oriented nature of the HiCAT software infrastructure. The goal of the design is to easily enable any code to run on HiCAT autonomously while monitoring safety concerns. Developers are still able to directly control hardware for quick-and-dirty troubleshooting, assuming they are present in the lab and can ensure safe operating conditions manually. Although, for formal experiments, a class gets created that inherits \verb|Experiment| and implements the abstract function \verb|experiment()|.  

The flow of every experiment initially checks all safety tests, and only if they all pass does it spin off a new process and start the experiment.  While the experiment is running on a separate process, the main process checks each safety test at given intervals using a semaphore-like design to go to sleep and wake up appropriately.

\subsection{Event-based Experiment Queuing}
Launcher scripts queue up a set of experiments to run sequentially. Within the launcher, a list of \verb|Experiment| subclasses is created to be iterated over and run. Each experiment is wrapped in a try-catch statement specifically catching the custom \verb|Safety Exception| which is triggered by the safety monitoring code. This \verb|Safety Exception| will cause the rest of the items in the queue to be skipped entirely and the testbed properly shutdown.  Other exceptions caused by anomalies or bugs in the code are also caught, although control will be gracefully passed back to the launcher which starts the next experiment. The concept and implementation of the \verb|Experiment| class and launcher will allow a CI system to safely and autonomously run tests on the hardware that are event-based and continue even when unexpected, or expected, failures occur.  

\begin{figure}
	\centering
	\includegraphics[scale=.5]{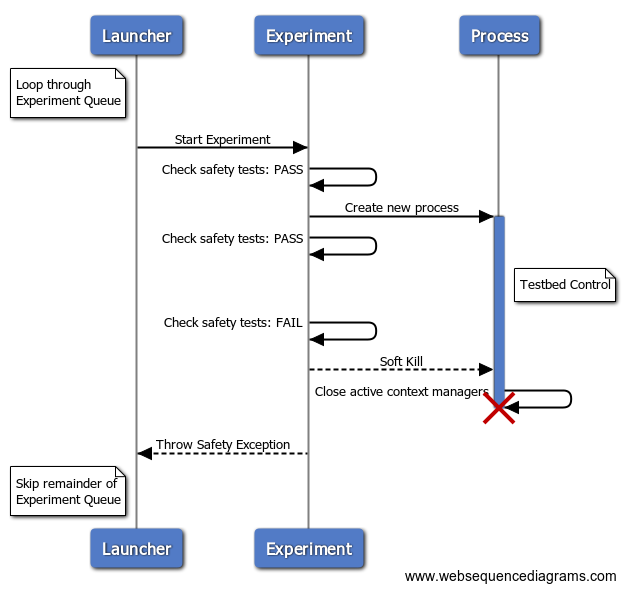} 
    \caption{Sequence diagram showing how the multiple processes are used to perform experiments and safety monitoring. The example illustrates when a safety test fails and shows how the system shuts down.}
    \label{fig:sequence_diagram}
\end{figure}

\subsection{Data Pipeline}
Some experiments require that images are processed in real-time in order to iterate and continue.  To account for this we developed a highly-customizable data pipeline that can be run on sets of FITS files or images held in memory. The pipeline averages images, removes bad pixels, performs image registration, and automatically populates the FITS header with metadata. If an experiment does not require a real-time data pipeline, it can be turned off and instead run later on a server with more processing resources. The data pipeline is configured by keyword arguments that can be passed into the \verb|Experiment| subclass.

\section{Simulation and Mocking}
\label{sec:Simulator}
The object-oriented hardware interface design presented allows for both mocking and simulation to be implemented in such a way that would allow a CI system to easily switch between actual hardware and simulation. Initially an external simulator was developed in Wolfram Mathematica for a specific experiment, although now a Python simulator is being created by subclassing the abstract hardware types. The new simulator makes use of the POPPY\cite{perrin2012} Python library to create a model of the optics of HiCAT, but then makes use of the existing code that operates the testbed. When the code makes a call to \verb|Camera::take_exposures()|, or \verb|DeformableMirrorController::apply_shape()|, it will run the simulated code instead. The simulated classes then create output products that follow the standard file format and closely resemble the images produced by HiCAT itself, which will allow for development and testing of new features to continue even when the testbed is offline.

Simulators are often as time-consuming to maintain and keep up-to-date as the system itself, especially with a project like HiCAT which has a constant flow of new optics and hardware to incorporate or exchange temporarily. Other forms of offline testing should be implemented for new features that the simulator does not yet support. This can be done by storing HiCAT output products and creating primitive subclasses of hardware types that simply return specific output products. This paradigm is called mocking. A Python mocking framework exists to facilitate more formal implementations and can be used to bypass an actual hardware API that cannot be initialized by an CI server when the testbed is unavailable. Mocking also promotes writing fast unit tests, which is integral for a CI system to be effective. Fast testing enables running these tests upon every new code change, and provides developers with instant feedback, catching regressions up front rather than after integration. 

At the time of writing, the Python simulator is near completion and will begin testing shortly. The mocking framework is still in early stages of implementation, and remains a proof-of-concept.    

\section{Continuous Integration with a Testbed}
\label{sec:Continuous Integration with a Testbed}

In the context of a Python software codebase, CI is primarily about unit and regression testing the code repository regularly. While there is a step to set up a new Python environment for each ``build", Python is an interpreted language and does not require a formal build or compilation step. Multiple unit test and continuous integration solutions were investigated, and the popular and well-supported pytest\cite{pytest} unit test framework and Jenkins continuous integration server were chosen.

\subsection{Unit and Regression Testing}
The reason a distinction between a unit test and regression test is made is to establish the difference between testing portions of code, and testing the functionality of the system. Traditionally, a unit test is written to test a certain module or function, aiming to have enough tests that cover the full codebase. However a regression test will test system level functionality and, for example, would run a set of experiments and compare the outputs. In the context of HiCAT, both are vital and will be implemented using a unit test framework, although they are fundamentally different and may be used at different steps in the software development workflow. 

Leveraging the object-oriented modular design of the HiCAT Python package, unit tests are more convenient to write, as opposed to the initial iteration of HiCAT control system (mostly notebooks, scripts, and pre-compiled executables). Such a system would require overhead before being able to test an individual portion of the system. As previously mentioned, fast tests are desired for CI, and the latter model ultimately tests the same parts of the system over and over, tending to result in longer test times and misleading test results. 

The majority of regression and unit tests will be written with the capability to be run with or without the testbed.  Fortunately, this can be implemented at an abstract level and does not require the developer to write multiple sets of tests for the same purpose.  Since the simulator and mock framework will take advantage of the object-oriented design, every test will start by taking a parameter which dictates which set of \verb|Instrument| subclasses it should run with.  Normally these are declared in the external configuration file, but it is overridden in this testing scenario.  For example, if the testbed is available, it will initialize the actual hardware interface classes, and specify \verb|ZwoCamera|, \verb|NewportMotorController|, \verb|BostonDmController|, etc.  Similarly, if new code is pushed and the developer is looking for a quick sanity check, they can run the tests by passing the \verb|simulator| or \verb|mock| parameter, and the appropriate classes will initialized. This capability will certainly prove useful when hardware, simulated, and mocked tests are correlated.

\subsection{Hardware Testing and Trending}
One of the interesting prospects of using CI on a testbed is to catch changes on a testbed that are unintentional, or slowly accumulated errors caused by mounts drifting, lasers changing in intensity, vibration and stability changes or any other of the countless variables involved with an optical testbed. Running tests regularly, perhaps even nightly, will allow trending that can show accumulations of testbed instability and allow problems to be identified early. The calibration suite for instance, is in the process of being converted to use pytest. Once wrapped as unit tests, the calibration activities can be routinely run and tracked as part of a Jenkins build, which will also provide custom reports. 

\subsection{Jenkins}
Jenkins is in the process of being set up to support the several different modes of testing required for HiCAT. It can be configured to have slave servers that are used as endpoints for running tests. The main server lives on a virtual machine, which is where the simulated and mock tests tied to a Jenkins build that gets triggered for every new code change in the repository are run. Within the GitFlow model, this means that when a developer pushes code to a feature branch, a test report will automatically be received, without having to use a personal machine to run tests. The PC that controls HiCAT will run a Jenkins slave server, and the Jenkins build for hardware will be started manually.

\section{Conclusion}
\label{sec:Conclusion}
Developing a hardware control system for a testbed presents a unique set of challenges in the context of implementing modern software engineering processes, namely CI.  Laying a foundation to prepare the software architecture for simulation and mocking is crucial to enabling such an environment.  This paper presents a path to design, develop and implement a hardware control infrastructure that can easily adapt to hardware changes, and well situates the code for unit testing and continuous integration even when the testbed is unavailable.

The HiCAT team replaced a disparate LabVIEW-based hardware control system with a unified Python-based software infrastructure. As part of this process, we adopted Git for version control and Jira for ticket tracking.  Gitflow was also adopted as a feature branching workflow, where each Jira ticket corresponds to a branch, and releases are intuitively stored on their own ``master" branch.  While developing the framework for unit testing, we immediately identified the need to be able to run tests when the testbed was unavailable. Our solution is currently in development, and leverages our object-oriented architecture to subclass simulated versions of hardware types.  A python based HiCAT simulator, which models the optical planes on HiCAT, is nearing completion and will allow most tests to run without access to the physical testbed. However, a mocking framework is also being used for new features that aren't yet incorporated into the simulator, and also provides a fast sanity check for every push to the repository (even branches). The unit tests are written using ``pytest" and will read a configuration parameter from an INI file to determine which classes to load, making it so that unit tests require no extra code to run in a simulator or mock mode. 

Safe and reliable control of our deformable mirrors drives much of our hardware control design. For this, proper resource management is implemented with a Python construct called ``context managers".  Using object-oriented design principles, a parent class called \verb|Instrument| was created which contains a generic implementation of a context manager.  Each hardware type then inherits this functionality which allows the Python keyword \verb|with| to be used to open connections to hardware and close them safely without the need to explicitly make calls to close or cleanup a hardware connection. All interfaces to hardware components on HiCAT inherit instrument, providing a standardized intuitive way for developers to script complex experiments without needing to understand the sometimes messy hardware communication code.    

Safe hardware control is one tenant to autonomous operations, another is to monitor humidity and power.  Parallel processing with Python's multiprocessing library is used in such a way that when unsafe conditions are detected, the main process can send a soft-kill signal that will trigger context managers active on another process to gracefully close and clean up any currently open hardware connections.  With this in place, the concept of an experiment was created in the form of an abstract class, giving any children this parallel processing functionality.  The \verb|Experiment| class checks all safety tests and spins off a separate process for the actual experiment to run on without being interrupted by safety checks. Several \verb|Experiment| objects can be iterated over by using a ``Launcher" which specifically watches for a safety exception. Other functional exceptions caused by bugs or other unintended problems will simply cause the next experiment in the queue to be started. This enables HiCAT to operate 24/7 and ensures that the testbed will be properly closed after experiments, whether they complete successfully or not.

We have developed a suite of automated calibration activities that allows us to recover from hardware changes.  The activities also store the results which can be used to track the testbed's stability over time.  These activities will be wrapped in the pytest framework in the near future and act as regression tests for the testbed.  Not only will these tests allow us to catch bugs in hardware, but it will also provide an early alert to unintentional drift or misalignment in the testbed.  All tests are in the process of being added to a CI solution called Jenkins, although the hardware tests will only be started manually.  The very same tests will be able to run on the simulated or mocked tests, and those will be run every time the repository is changed.  The Jenkins CI solution is currently a prototype, with a slave server running on the PC controlling HiCAT and the main server running on a virtual server.  

Overall, the time saved by creating a software infrastructure that allows the testbed to be run on nights and weekends has been priceless.  Soon the team will be completing the CI solution presented in this paper, allowing us to develop and test when the testbed is unavailable, trend and catch drifts in testbed alignment, and write single tests do both of these.  Enabling CI for HiCAT will provide even more savings in development time by enabling developers to write reliable code, and still get valuable feedback on new feature development no matter the state of HiCAT.

\acknowledgments 
This work is supported in part by the National Aeronautics and Space Administration under Grants NNX12AG05G and NNX14AD33G issued through the Astrophysics Research and Analysis (APRA) program (PI: R. Soummer) and by the STScI Director's Discretionary Research Fund.

\bibliography{bibliotheque}

\begin{thebibliography}{10}

\bibitem{2016SPIE.9904E..3CL}
{Leboulleux}, L., {N'Diaye}, M., {Riggs}, A.~J.~E., {Egron}, S., {Mazoyer}, J.,
  {Pueyo}, L., {Choquet}, E., {Perrin}, M.~D., {Kasdin}, J., {Sauvage}, J.-F.,
  {Fusco}, T., and {Soummer}, R., ``{High-contrast imager for Complex Aperture
  Telescopes (HiCAT). 4. Status and wavefront control development},'' in [{\em
  Space Telescopes and Instrumentation 2016: Optical, Infrared, and Millimeter
  Wave}{\nolinebreak\hspace{0.1em}]},  {\em \procspie} {\bf 9904},  99043C
  (July 2016).

\bibitem{2017SPIE10562E..2ZL}
{Leboulleux}, L., {N'Diaye}, M., {Mazoyer}, J., {Pueyo}, L., {Perrin}, M.,
  {Egron}, S., {Choquet}, E., {Sauvage}, J.-F., {Fusco}, T., and {Soummer}, R.,
  ``{Comparison of wavefront control algorithms and first results on the
  high-contrast imager for complex aperture telescopes (hicat) testbed},'' in
  [{\em Society of Photo-Optical Instrumentation Engineers (SPIE) Conference
  Series}{\nolinebreak\hspace{0.1em}]},  {\em Society of Photo-Optical
  Instrumentation Engineers (SPIE) Conference Series} {\bf 10562},  105622Z
  (Sept. 2017).

\bibitem{ndiaye2015AAS}
{N'Diaye}, M., {Choquet}, E., {Carlotti}, A., {Pueyo}, L., {Egron}, S.,
  {Leboulleux}, L., {Levecq}, O., {Perrin}, M.~D., {Wallace}, J.~K., {Long},
  C., {Lajoie}, R., {Lajoie}, C.-P., {Eldorado Riggs}, A.~J., {Zimmerman},
  N.~T., {Groff}, T.~D., {Kasdin}, N.~J., {Vanderbei}, R.~J., {Mawet}, D.,
  {Macintosh}, B., {Shaklan}, S., and {Soummer}, R., ``{High-contrast imager
  for Complex Aperture Telescopes (HiCAT): APLC/shaped-pupil hybrid coronagraph
  designs},'' in [{\em American Astronomical Society Meeting
  Abstracts}{\nolinebreak\hspace{0.1em}]},  {\em American Astronomical Society
  Meeting Abstracts} {\bf 225},  258.09 (Jan. 2015).

\bibitem{ndiaye2014SPIE_hicat2}
{N'Diaye}, M., {Choquet}, E., {Egron}, S., {Pueyo}, L., {Leboulleux}, L.,
  {Levecq}, O., {Perrin}, M.~D., {Elliot}, E., {Wallace}, J.~K., {Hugot}, E.,
  {Marcos}, M., {Ferrari}, M., {Long}, C.~A., {Anderson}, R., {DiFelice}, A.,
  and {Soummer}, R., ``{High-contrast Imager for Complex Aperture Telescopes
  (HICAT): II. Design overview and first light results},'' in [{\em Society of
  Photo-Optical Instrumentation Engineers (SPIE) Conference
  Series}{\nolinebreak\hspace{0.1em}]},  {\em Society of Photo-Optical
  Instrumentation Engineers (SPIE) Conference Series} {\bf 9143},  27 (Aug.
  2014).

\bibitem{ndiaye2013SPIE}
{N'Diaye}, M., {Choquet}, E., {Pueyo}, L., {Elliot}, E., {Perrin}, M.~D.,
  {Wallace}, J.~K., {Groff}, T., {Carlotti}, A., {Mawet}, D., {Sheckells}, M.,
  {Shaklan}, S., {Macintosh}, B., {Kasdin}, N.~J., and {Soummer}, R.,
  ``{High-contrast imager for complex aperture telescopes (HiCAT): 1. testbed
  design},'' in [{\em Techniques and Instrumentation for Detection of
  Exoplanets VI}{\nolinebreak\hspace{0.1em}]},  {\em \procspie} {\bf 8864},
  88641K (Sept. 2013).

\bibitem{ndiaye2014SPIE_aplc}
{N'Diaye}, M., {Pueyo}, L., {Soummer}, R., and {Carlotti}, A., ``{Apodized
  Pupil Lyot Coronagraphs: development of designs with reduced IWA and
  robustness to low-order aberrations},'' in [{\em Society of Photo-Optical
  Instrumentation Engineers (SPIE) Conference
  Series}{\nolinebreak\hspace{0.1em}]},  {\em Society of Photo-Optical
  Instrumentation Engineers (SPIE) Conference Series} {\bf 9143},  4 (Aug.
  2014).

\bibitem{soummer2018}
{Soummer}, R., {Brady}, G.~R., {Brooks}, K., {Comeau}, T., {Dillon}, T.,
  {Choquet}, {\'E}., {Egron}, S., {Gontorum}, R., {Hagopian}, J., {Laginja},
  I., {Leboulleux}, L., {Perrin}, M.~D., P., P., {Pueyo}, L., {Mazoyer}, J.,
  {N'Diaye}, M., {Shiri}, R., {Sivaramakrishnan}, A., {St. Laurent}, K.,
  {Valenzuela}, A.-M., and {Zimmerman}, N., ``{High-contrast imager for complex
  aperture telescopes (HiCAT): 5. rst results with segmented-aperture
  coronagraph and wavefront control},'' in [{\em UV/Optical/IR Space Telescopes
  and Instruments: Innovative Technologies and Concepts
  VIII}{\nolinebreak\hspace{0.1em}]},  {\em SPIE paper 10698-59 in these
  proceedings} (2018).

\bibitem{jenkins}
``Jenkins.'' \url{https://jenkins.io/}.
\newblock Online; accessed June 14, 2018].

\bibitem{numpy}
Oliphant, T., ``A guide to numpy,,'' (2006).

\bibitem{scipy}
Jones, E., Oliphant, T., Peterson, P., et~al., ``{SciPy}: Open source
  scientific tools for {Python},'' (2001--).
\newblock [Online; accessed June 14, 2018]].

\bibitem{matlab_to_python}
Sievert, S., ``Stepping from matlab to python,'' (2015).
\newblock [Online; accessed June 14, 2018]].

\bibitem{laginja2018}
{Laginja}, I., {Brady}, G., {Soummer}, R., {Egron}, S., {Lajoie}, C.,
  {Bonnefois}, A., {Michau}, V., {Choquet}, E., {Ferrari}, M., L., L.,
  {Levecq}, L., {Mazoyer}, J., {N'Diaye}, M., {Perrin}, M., {Petrone}, P.,
  {Pueyo}, L., and {Sivaramakrishnan}, A., ``{James Webb Space telescope
  optical simulation testbed V: comparison of wide-field phase retrieval
  techniques },'' in [{\em UV/Optical/IR Space Telescopes and Instruments:
  Innovative Technologies and Concepts VIII}{\nolinebreak\hspace{0.1em}]},
  {\em SPIE paper 10698-126 in these proceedings} (2018).

\bibitem{brady2018}
{Brady}, G., {Moriarty}, C., {Petrone}, P., {Laginja}, I., {Brooks}, K.,
  {Comeau}, T., {Leboulleux}, L., and {Soummer}, R., ``{Phase-retrieval-based
  wavefront metrology for high contrast coronagraphy},'' in [{\em UV/Optical/IR
  Space Telescopes and Instruments: Innovative Technologies and Concepts
  VIII}{\nolinebreak\hspace{0.1em}]},  {\em SPIE paper 10698-235 in these
  proceedings} (2018).

\bibitem{gil}
``Global interpreter lock.''
  \url{https://wiki.python.org/moin/GlobalInterpreterLock}.
\newblock Online; accessed June 14, 2018].

\bibitem{perrin2012}
{Perrin}, M.~D., {Soummer}, R., {Elliott}, E.~M., {Lallo}, M.~D., and
  {Sivaramakrishnan}, A., ``{Simulating point spread functions for the James
  Webb Space Telescope with WebbPSF},'' in [{\em Space Telescopes and
  Instrumentation 2012: Optical, Infrared, and Millimeter
  Wave}{\nolinebreak\hspace{0.1em}]},  {\em \procspie} {\bf 8442},  84423D
  (Sept. 2012).

\bibitem{pytest}
holger krekel and pytest-dev team, ``pytest,'' (2015).
\newblock [Online; accessed June 14, 2018].

\end{thebibliography}
\bibliographystyle{spiebib}

\end{document}